# 3-List Colouring Permutation Graphs

Jessica Enright      Lorna Stewart

**Abstract**

3-list colouring is an NP-complete decision problem. It is hard even on planar bipartite graphs. We give a polynomial-time algorithm for solving 3-list colouring on permutation graphs.

## 1  Introduction

In the vertex colouring problem, we try to assign each vertex in a graph a colour such that no two adjacent vertices are assigned the same colour using the minimum number of colours.

In the vertex list colouring problem, each vertex has a list of colours, and we try to assign each vertex a colour from its list such that no two vertices are assigned the same colour. Determining if this is possible is a decision problem and is NP-complete, as it is a generalization of vertex colouring [3]. List colouring remains hard even on interval graphs [2].

List colouring with fixed colour bound of 3 is a generalization of 3-vertex colouring, and so is NP-complete. It remains NP-complete on planar bipartite graphs [4], but is solvable in polynomial time on graphs of fixed treewidth [2]. We give a polynomial-time algorithm for solving list colouring with fixed colour bound three on a class of intersection graphs that does not have fixed treewidth.

Permutation graphs are exactly comparability cocomparability graphs - the graphs that admit transitive orientations of both their edges and their nonedges. In this paper we give a polynomial algorithm for list colouring with a fixed colour bound of 3 on permutation graphs.

Our algorithm uses the layers of a breadth-first search rooted at a particular vertex in what we call a multi-chain ordering. This ordering is closely related to the strong ordering used by Heggernes et al. [1] to compute the bandwidth of bipartite permutation graphs in polynomial time. Our ordering, which applies to the larger class of permutation graphs, is expressed in terms of the layers of a breadth first search. This sort of ordering gives insight into the structure of permutation graphs, and may lead to to algorithms for other problems on permutation graphs.

## 2  Definitions and Preliminaries

A graph $G = (V, E)$ is a tuple of vertex set $V$ and edge set $E$ composed of subsets of $V$ of size two. All graphs that we consider are connected, finite, simple, and loopless. A directed graph $G = (V, E)$ is a tuple of vertex set $V$ and edge set $E$ composed of ordered pairs of $V$.

A transitive orientation of the edges of a graph is an orientation such that the presence of edge $(u \to v)$ and $(v \to w)$ implies edge $(u \to w)$. A comparability graph is a graph that admits a transitive orientation of its edges. A cocomparability graph is a graph that admits a transitive orientation of its nonedges.

Let $G = (V, E)$ be a graph. A *list mapping* of $G$ is a mapping that assigns to each vertex in $G$ a list of colours. A colouring of $G$ that obeys a list mapping $\mathcal{P}$ of $G$ is a colouring such that every vertex is assigned a colour that is in that vertex's list in $\mathcal{P}$. A 3-list colouring is a proper list colouring in which at most three colours are used.

We might sometimes say that a list mapping *precolours* a vertex. By this we mean that the list mapping assigns a list with only a single colour to that vertex.

## 2.1 Multi-chain ordering

Let $G = (V, E)$ be a graph and $\mathcal{L} = [L_0...L_k]$ be the layers of a breadth-first traversal of $G$ with vertex $v_0$ as the starting point. We call $\mathcal{L}$ a *multi-chain ordering* of $G$ if for every two vertices $u, v$ in layer $L_i$ the neighbourhood of $u$ in $L_{i-1}$ is a subset of the neighbourhood of $v$ in $L_{i-1}$ (or vice versa) and the neighbourhood of $u$ in $L_{i+1}$ is a subset of the neighbourhood of $v$ in $L_{i+1}$ (or vice versa).

**Lemma 1.** *Let $\overrightarrow{G} = (V, \overrightarrow{E})$ be a transitive orientation of a comparability graph $G = (V, E)$ in which $v_0$ is a source or a sink, and $[L_0...L_k]$ be the layers of a breadth first search traversal of $G$ starting at $v_0$. For every two consecutive layers $L_i, L_{i+1}$ for $0 \le i < k$, either all edges in $\overrightarrow{E}$ between vertices of $L_i$ and $L_{i+1}$ are directed toward $L_i$ or all edges in $\overrightarrow{E}$ between vertices of $L_i$ and $L_{i+1}$ in are directed toward $L_{i+1}$.*

*Proof.* This follows from the fact that $v_0$ is a source or sink in $\overrightarrow{G}$, and the observation that there are no edges between nonconsecutive layers of a breadth first search. □

**Lemma 2.** *Let $\overrightarrow{G} = (V, \overrightarrow{E})$ be a transitive orientation of the complement of a comparability graph $G = (V, E)$ in which $v_0$ is a sink and $\mathcal{L} = [L_0...L_k]$ are the layers of a breadth-first search traversal of $G$ starting at $v_0$. Then for every pair of layers $L_i, L_j$ where $0 \le i < j \le k$ all nonedges between $L_i$ and $L_j$ are directed toward $L_i$.*

*Proof.* We proceed by induction. First, consider the level $L_0$. Because the only vertex on $L_0$ is a sink in $\overline{G}$, all nonedges between another level and the vertex on level $L_0$ are directed toward $L_0$.

Assume that for every layer $L_h$ such that $h \le i$, all nonedges between $L_h$ and a layer of index greater than $h$ are directed toward $L_h$. Then consider $L_{i+1}$.

Now we continue by contradiction. Assume that there is a nonedge between vertex $v_{i+1}$ in $L_{i+1}$ and vertex $v_j$ in some layer $L_j$ where $j > i$ that is directed toward $L_j$. Let $v_i$ be a neighbour of $v_{i+1}$ in layer $L_i$. Because the layers are produced by a breadth first traversal, there is a nonedge between $v_j$ and $v_i$, and by the inductive assumption, it is directed toward $v_i$. Then we have in $G$ a nonedge directed from $v_{i+1}$ to $v_j$ and from $v_j$ to $v_i$, but no nonedge between $v_i$ and $v_{i+1}$, a contradiction. □

**Lemma 3.** *Let $G = (V, E)$ be a permutation graph and let $\overrightarrow{G}$ be a transitive orientation of $G$ in which $v_0$ is a source or a sink, and $\overrightarrow{\overline{G}}$ a transitive orientation of the complement of $G$ in which $v_0$ is a sink. Let $\mathcal{L} = [L_0...L_k]$ be the layers of a breadth-first search traversal of $G$ rooted at $v_0$. Then $\mathcal{L}$ is a multi-chain ordering.*

*Proof.* Let $u, v$ be two vertices on $L_i$. We consider two cases: if $u, v$ are adjacent, and if they are not adjacent.

We proceed by contradiction. Assume that $u, v$ are adjacent and that there are vertices $x, y$ in $L_{i+1}$ such that $x$ is adjacent to $u$ but not $v$ and $y$ is adjacent to $v$ but not $u$. By Lemma 1, the edges between $u$ and $x$ and between $v$ and $y$ are either both directed toward $L_i$, or both directed toward $L_{i+1}$. In either case there is no transitive orientation of the edge between $u$ and $v$, a contradiction. Similarly, if $x, y$ are in $L_{i-1}$ by Lemma 1, the edges are either both directed toward $L_i$, or both directed toward $L_{i+1}$. In either case there is no transitive orientation of the edge between $u$ and $v$, a contradiction.

Now assume that $u, v$ are not adjacent and that there are vertices $x, y$ in $L_{i+1}$ such that $x$ is adjacent to $u$ but not $v$ and $y$ is adjacent to $v$ but not $u$. By Lemma 2, these nonedges are directed as $(x \to v)$ and $(y \to u)$. Then there is no transitive orientation of the nonedge between $u$ and $v$, a contradiction. Similarly, if $x, y$ are in $L_{i-1}$ then by Lemma 2, these nonedges are directed as $(v \to x)$ and $(u \to y)$. Then there is no transitive orientation of the nonedge between $u$ and $v$, a contradiction.

We have shown that whether or not two vertices $u, v$ on layer $L_i$ are adjacent, the neighbourhood of one in $L_{i-1}$ is a subset of the neighbourhood of the other in $L_{i-1}$, and the neighbourhood of one in $L_{i+1}$ is a subset of the neighbourhood of the other in $L_{i+1}$. □

**Theorem 4.** *Every permutation graph has a multi-chain ordering.*

*Proof.* This follows from the comparability cocomparability orderings of permutation graphs and the previous three lemmas. □

But also observe that:

**Observation 5.** *Not every graph with a multi-chain ordering is a permutation graph.*

What does this neighbourhood containment mean for our list colouring? In each layer there is a vertex with a maximal neighbourhood in the previous layer, and a vertex with a maximal neighbourhood in the following layer. Because each vertex in a layer has at least one neighbour in the previous layer, this means that there is a vertex in each layer adjacent to all vertices in the next layer. Therefore:

**Observation 6.** *No single layer in a multi-chain ordering can have vertices of more than two colours in it in a valid 3-list-colouring.*

Also observe that because there is a vertex in each layer adjacent to all vertices in the next layer that there are no triangles on any layer. Then we can deal with each layer as a collection of connected bipartite components.

Let $G = (V, E)$ be a graph with multi-chain ordering $\mathcal{L} = [L_0...L_k]$, and list mapping $\mathcal{P}$. A *good colour assignment* $Q$ is a relation between the three colours and the layers in $\mathcal{L}$ such that each layer is assigned exactly two colours, two adjacent layers are not assigned the same two colours, and if there is a vertex on layer $L_i$ that is precoloured $C_x$ by $\mathcal{P}$, then $C_x$ is one of the colours assigned to $L_i$ by $Q$. We use $Q(i)$ to denote the colours assigned by $Q$ to layer $i$.

A list colouring $\mathcal{C}$ of graph $G$ *admits* $Q$ if every vertex is coloured by $\mathcal{C}$ with a colour assigned to its layer by $Q$.

A layer is precoloured with a colour by $\mathcal{P}$ if there is at least one vertex on layer $L_i$ that is precoloured with that colour by $\mathcal{P}$.

Let vertex $v_i$ be on layer $L_i$ of a multi-chain ordering. Vertex $v_i$ is *quasi-precoloured* by colour assignment $Q$ and list mapping $\mathcal{P}$ if any of the following are true:

- There is a single colour in the intersection of $\mathcal{P}(v_i)$ and $Q(i)$.

- $v_i$ is adjacent to two vertices that are on opposite sides of a connected bipartite component of one of $G[L_{i-1}]$ or $G[L_{i+1}]$

- $v_i$ is in a connected bipartite component of $G[L_i]$ such that at least one other vertex in that connected bipartite component is quasi-precoloured

Vertices $v_i$ in $L_i$, and $v_{i+1}$ in $L_{i+1}$ are *almost-adjacent* if $v_{i+1}$ is part of a bipartite component $C_j$ with bipartition $B_1, B_2$ where $v_{i+1}$ is in $B_2$ and $v_i$ is adjacent to at least one vertex in $B_2$. Two vertices are *almost-neighbours* if they are almost-adjacent. An *almost-path* $P = [v_1...v_k]$ is a sequence of vertices such that for $1 < i < k$ vertex $v_i$ is almost-adjacent to vertices $v_{i-1}$ and $v_{i+1}$.

A *quasi-bad chain* $P = [v_h...v_l]$ is a sequence of vertices in layers $L_h...L_l$ such that:

- Each vertex is in a different layer, proceeding consecutively through $\mathcal{L}$ and

- Vertices $v_i, v_{i+1}$ are either adjacent or almost-adjacent and

- Each layer $L_i$ for $h < i < l$ is assigned two colours $C_x, C_y$ by $Q$ such that $C_x$ is assigned to $L_{i-1}$ and $C_y$ is assigned to $L_{i+1}$ and

- $v_h$ is quasi-precoloured with the colour assigned to $L_{h+1}$ but not $L_{h+2}$, and

- $v_l$ is quasi-precoloured with the colour assigned to $L_{l-1}$ but not $L_{l-2}$.

Let $Q$ be a good assignment of colours to multi-chain ordering $\mathcal{L}$. Let $P = [v_h...v_l]$ be a quasi-bad chain, given $Q$. Observe that:

**Observation 7.** *$Q$ does not assign the same colour to three consecutive layers that contain vertices of $P$, nor does it assign the same colour pairs to two consecutive layers.*

Let $L_l$ be the highest-indexed layer in $\mathcal{L}$ that contains a vertex of quasi-bad chain $P$. We say that $P$ is an *N2* quasi-bad chain if for every other quasi-bad chain $P'$ with respect to $Q$, the highest-indexed layer containing a vertex of $P'$ has index at most $h$.

Given a good colour assignment $Q$ to multi-chain ordering $\mathcal{L}$ of $G = (V, E)$ with list mapping $\mathcal{P}$, how can we produce a 3-list colouring of the vertices of $G$ that is consistent with $Q$ and $\mathcal{P}$?

We claim that we can use a simple greedy approach as given in Algorithm 1. This algorithm traverses the layers of the multi-chain ordering from smallest to largest index and colours vertices in each layer first if they are quasi-precoloured. If a vertex is not quasi-precoloured, the algorithm tries to colour the vertex with the colour its layer does not share with the next layer. As a last resort, it colours a vertex with the colour its layer shares with the next layer. The algorithm handles all the vertices of a connected bipartite components of the layer as a group.

**Lemma 8.** *Let $G = (V, E)$ be a graph with chain ordering $\mathcal{L} = [L_0...L_k]$, list mapping $\mathcal{P}$, and let $Q$ be a good assignment of colours to layers in $\mathcal{L}$. If there is no 3-list colouring of $G$ that admits $Q$, then there exists a quasi-bad chain $P = [v_h...v_l]$ within $\mathcal{L}$ with respect to colour assignment $Q$.*

*Proof.* If there is no 3-list colouring of $G$ that admits $Q$, then Algorithm 1 will fail.

We first show that if the conservative colouring algorithm fails in the component colouring section, then there is a quasi-bad chain $P = [v_h...v_l]$ within $\mathcal{L}$ with respect to colour assignment $Q$. Assume that the component colouring algorithm fails. Then let $C$ be the component on layer $L_i$ on which the section fails and $C_x$ and $C_y$ be the colours assigned to $L_i$ by $Q$. There are no vertices in $C$ that are quasi-precoloured by $\mathcal{P}$.

Then either a single vertex in $C$ has neighbours outside of $L$ of both colours $C_x, C_y$, or vertices on both sides of the bipartition of $C$ have neighbours of one of colour $C_x$ or $C_y$. We consider these three cases.

If a single vertex in $C$ has neighbours outside of $L$ of both colours $C_x, C_y$, the neighbour $v_{i+1}$ of colour $C_y$ is in layer $L_{i+1}$ and the neighbour $v_{i-1}$ of colour $C_x$ is in layer $L_{i-1}$. Because the execution of Algorithm 1 has not yet reached $L_{i+1}$, vertex $v_{i+1}$ must be quasi-precoloured.

Vertex $v_i$ is a vertex that is not quasi-precoloured, is coloured by the conservative algorithm with the colour layer $L_i$ shares with layer $L_{i+1}$ and there is an almost-path between $v_{i-1}$ and $v_i$ such that every vertex on the path is not precoloured, and is coloured by the conservative algorithm with the colour its layer shares with the next layer.

Let $v_j$ on level $L_j$ be a vertex with those properties such that $j$ is minimal. Then $v_j$ is only coloured with the colour $L_j$ shares with $L_{j+1}$ because it has a neighbour of the colour $L_j$ does not share with $L_{j+1}$. Then, by the minimality of $j$, this neighbour must be quasi-precoloured. We refer to it as $v_{j-1}$. The path from $v_{j-1}$ to $v_{i+1}$ that contains no other quasi-precoloured vertices is by definition a quasi-bad chain.

If vertices on both sides of the bipartition of $C$ have neighbours of colour $C_y$, then at least one of those neighbours is adjacent to vertices on both sides of the bipartition of $C$, and so it is quasi-precoloured by $\mathcal{P}$ with colour $C_z$, a contradiction to it being coloured $C_y$.

Similarly, if vertices on both sides of the bipartition of $C$ have neighbours of colour $C_x$, then we can derive a contradiction.

We now proceed assuming that the Conservative Colouring Algorithm fails at some point other than while executing the component colouring section. Let $v_i$ be the first vertex encountered by Algorithm 1 that cannot be coloured and $C_x, C_y$ be the colours assigned to $L_i$.

Vertex $v_i$ has at least one almost-neighbour that is coloured each of $C_x, C_y$. Because the failure is outside the component colouring section, neither of these coloured almost-neighbours are in layer $L_i$.

Since $L_i$ shares at most one colour with $L_{i-1}$ and at most one with $L_{i+1}$, one of these neighbours is on level $L_{i-1}$ and one (we will call it $v_{i+1}$) on $L_{i+1}$. Because of the execution of Algorithm 1, $v_{i+1}$ is quasi-precoloured.

Vertex $v_i$ is a vertex that is not quasi-precoloured, is coloured by the conservative algorithm with the colour layer $L_i$ shares with layer $L_{i+1}$ and there is a almost-path between $v_{i-1}$ and $v_i$ such that every vertex on the path is not precoloured, and is coloured by the conservative algorithm with the colour its layer shares with the next layer.

Let $v_j$ on level $L_j$ be the lowest-numbered vertex with that property. Then $v_j$ is only coloured with the colour $L_j$ shares with $L_{j+1}$ because it has an almost-neighbour of the colour $L_j$ does not share with $L_{j+1}$. Then, by the minimality of $j$, this almost-neighbour must be quasi-precoloured. We refer to it as $v_{j-1}$. The path from $v_{j-1}$ to $v_{i+1}$ that contains no other quasi-precoloured vertices is by definition a quasi-bad chain. $\square$

Then from the previous lemma and the fact that a quasi-bad chain prevents any 3-list colouring it follows that:

**Lemma 9.** *Let $G = (V, E)$ be a graph with chain ordering $\mathcal{L} = [L_0...L_k]$, and list mapping $\mathcal{P}$. Let $Q$ be a good assignment of colours to layers in $\mathcal{L}$ that obeys the allowable pairs. There is a 3-list colouring of $G$ that admits $Q$ if and only if there is no quasi-bad chain within $\mathcal{L}$, with respect to $Q$.*

## 3 Allowable Colour Array

Our approach to finding a 3-list colouring for a graph $G = (V, E)$ with multi-chain ordering $\mathcal{L} = [L_0...L_k]$ is based on finding a good colour assignment $Q$ of two colours to each layer in $\mathcal{L}$ such that there is a 3-list colouring of $G$ in which each vertex on a layer is coloured with one of the two colours assigned to that layer.

To do this we use an *allowable colour array*. For a graph $G = (V, E)$ with multi-chain ordering $\mathcal{L} = [L_0...L_k]$ an allowable array $\mathcal{A}$ will be an array of length $k$. Each entry $\mathcal{A}[i]$ will contain a list of the pairs of colour assignments that are allowable for $L_i, L_{i+1}$.

What do we mean by allowable? We mean that they have not been precluded by any of a number of rules we will develop. For example, an assignment of colours $1, 2$ is not allowable to a layer that has on it a vertex precoloured with the colour 3. The remainder of this section is concerned with formally describing the allowable array, and a polynomial-time algorithm for computing it given a graph with a list mapping.

**Algorithm 1** Conservative Colouring Algorithm. Input: Graph $G$, list mapping $\mathcal{P}$, good colour assignment $Q$, multi-chain ordering $\mathcal{L}$, integer start, integer end. Returns: false if forced into an incorrect colouring, true otherwise.

---

1: **for all** $i$ from *start* to *end* **do**
2:     notBelow ← the colour assigned to $L_i$ but not $L_{i+1}$ by $Q$
3:     withBelow ← the colour assigned to both $L_i$ and $L_{i+1}$ by $Q$
4:     **for all** uncoloured $v_j \in L_i$ **do**
5:         **if** $v_j$ is in a connected bipartite component of more than two vertices in $L_i$ **then**
6:             COMMENT: Beginning of Component Colouring Section
7:             $C \leftarrow$ component in $L_i$ that contains $v$
8:             $B_1 \leftarrow$ one side of bipartition of $C$
9:             $B_2 \leftarrow$ other side of bipartition of $C$
10:            notBelow ← the colour assigned to $L_i$ but not $L_{i+1}$ by $Q$
11:            withBelow ← the colour assigned to both $L_i$ and $L_{i+1}$ by $Q$
12:            **if** a vertex in $B_1$ has neighbour of colour notBelow **then**
13:                colour all vertices in $B_1$ with colour withBelow
14:                colour all vertices in $B_2$ with colour notBelow
15:            **else if** a vertex in $B_2$ has neighbour of colour notBelow **then**
16:                colour all vertices in $B_1$ with colour notBelow
17:                colour all vertices in $B_2$ with colour withBelow
18:            **else**
19:                $v_{B_1} \leftarrow$ member of $B_1$ with maximal neighbourhood in $L_{i+1}$
20:                $v_{B_2} \leftarrow$ member of $B_2$ with maximal neighbourhood in $L_{i+1}$
21:                **if** $(L_{i+1} \cap N(v_{B_1})) \subset (L_{i+1} \cap N(v_{B_2}))$ **then**
22:                   colour all vertices in $B_1$ with colour withBelow
23:                   colour all vertices in $B_2$ with colour notBelow
24:                **else**
25:                   colour all vertices in $B_1$ with colour notBelow
26:                   colour all vertices in $B_2$ with colour withBelow
27:                **end if**
28:            **end if**
29:             $C_{B_1} \leftarrow$ colour assigned to vertices in $B_1$
30:             $C_{B_2} \leftarrow$ colour assigned to vertices in $B_2$
31:             **if** any vertex in $B_1$ has a neighbour of colour $C_{B_1}$ **then**
32:                **return** false
33:             **end if**
34:             **if** any vertex in $B_2$ has a neighbour of colour $C_{B_2}$ **then**
35:                **return** false
36:             **end if**
37:             COMMENT: End of Component Colouring Section
38:         **else if** $v_j$ has a neighbour of colour notBelow **then**
39:             colour $v_j$ with withBelow
40:             **if** $v_j$ has a neighbour of colour withBelow **then**
41:                **return** false
42:             **end if**
43:         **else**
44:             colour $v_j$ with notBelow
45:         **end if**
46:     **end for**
47: **end for**
48: **return** true

Let $\mathcal{A}$ be an array of length $k$ such that each entry $\mathcal{A}[i]$ contains a list of pairs of colour assignments. Each colour assignment must consist of two distinct colours. Let $Q$ be a colour assignment of two colours to each layer in $\mathcal{L}$ such that for every two adjacent layers $L_i, L_{i+1}$, the colour assignments made by $Q$ to $L_i, L_{i+1}$ are listed in $\mathcal{A}[i]$.

Let $L_i, L_{i+1}$ be adjacent layers that are assigned $C_x, C_y$ and $C_y, C_z$, respectively, by $Q$. We say that $L_i$ is *adjustable* with respect to $\mathcal{A}$ if $C_x, C_z$ and $C_y, C_z$ are listed as a pair in $\mathcal{A}[i]$.

Let $P = [v_h...v_l]$ be a quasi-bad chain, and $L_h...L_l$ the layers that contain members of $P$. We say that the colour assignments by $Q$ to $L_l, L_{l-1}$ are *quasi-bad forcing* $P$ with respect to $\mathcal{A}$ if no colour assignment to a layer $L_i$ where $h \leq i < l-1$ is adjustable with respect to $\mathcal{A}$.

Let $\mathcal{A}$ be an array of length $k$ such that each entry $\mathcal{A}[i]$ contains a list of pairs of colour assignments, each consisting of two distinct colours. $\mathcal{A}$ is an *allowable array* for $G = (V, E), \mathcal{L}, \mathcal{P}$ if there is no pair of colour assignments $C_x, C_y$ and $C_y, C_z$ to layers $L_i, L_{i+1}$ such that one of the following is true:

- there is a vertex on $L_i$ precoloured by $\mathcal{P}$ with colour $C_z$, or a vertex on $L_{i+1}$ precoloured by $\mathcal{P}$ with colour $C_x$,

- assigning these colours to these layers results in a quasi-precolouring such that two adjacent vertices are given the same colour,

- there is no pair of colour assignments for $L_{i-1}, L_i$ that assigns $C_x, C_y$ to $L_i$,

- there is no pair of colour assignments for $L_{i+1}, L_{i+2}$ that assigns $C_x, C_z$ to $L_{i+1}$ ,

- the colour assignments $C_x, C_y$ and $C_y, C_z$ to layers $L_i, L_{i+1}$ are quasi-bad forcing.

We say that a colour assignment $Q$ to layers $\mathcal{L}$ *obeys* an allowable array $\mathcal{A}$ if for every two adjacent layers $L_i, L_{i+1}$, the colour assignments made by $Q$ to $L_i, L_{i+1}$ are listed as a possibility in $\mathcal{A}[i]$.

**Lemma 10.** *If Algorithm 2 removes a pair of colour assignments from $\mathcal{A}$ then there is no colouring of $G$ that is consistent with those removed colour assignments. Algorithm 2 produces the allowable array of the input graph $G$, chain ordering $\mathcal{L}$ and list mapping $\mathcal{P}$.*

*Proof.* We will argue this inductively.

We will show that if we remove an allowable pair from the listing during execution of Algorithm 2, then there is no colouring of $G$ that admits $\mathcal{P}$ in which that pair of colour assignments were assigned to their specified layers.

Consider the first allowable pair that we remove. Let the pair removed be $C_x, C_y$ assigned to layer $L_i$, and $C_y, C_z$ assigned to $L_{i+1}$. There are four cases.

- There is a vertex precoloured by $\mathcal{P}$ with $C_z$ on $L_i$, or a vertex precoloured by $\mathcal{P}$ with $C_x$ on $L_{i+1}$.

- Assigning $C_x, C_y$ to layer $L_i$, and $C_y, C_z$ to layer $L_{i+1}$ results in two adjacent vertices being quasi-precoloured the same colour.

- There is no allowable pair for $L_{i-1}, L_i$ in which $L_i$ is assigned $C_x, C_y$, or there is no allowable pair for $L_{i+1}, L_{i+2}$ in which $L_{i+1}$ is assigned $C_y, C_z$.

- Assigning $C_x, C_y$ to layer $L_i$, and $C_y, C_z$ to layer $L_{i+1}$ quasi-bad forces a bad chain $P = [v_h...v_{i+1}]$.

Because we call Algorithm 2 with an array that initially contains all possible colour assignments for every pair of layers, no colour assignments can be quasi-bad forcing, so every layer is adjustable. Similarly, with an initially-full array, neither of the if statements on lines 11 or 15 will evaluate to true.

Therefore, only the first two cases can result in the first removal of a colour assignment pair. In either case, there is no colouring of $G$ that admits $\mathcal{P}$ such that $L_i$ and $L_{i+1}$ are assigned $C_x, C_y$ and $C_y, C_z$.

Now, assume that for each of the first $k$ allowable pairs removed by Algorithm 2, there is no colouring of $G$ that admits $Q$ such that that allowable pair is assigned to the specified layers.

Then let the $(k+1)^{th}$ pair removed be $C_x, C_y$ assigned to layer $L_i$, and $C_y, C_z$ assigned to $L_{i+1}$.

As in the base case, there are four cases.

- There is a vertex precoloured by $\mathcal{P}$ with $C_z$ on $L_i$, or a vertex precoloured by $\mathcal{P}$ with $C_x$ on $L_{i+1}$.

- Assigning $C_x, C_y$ to layer $L_i$, and $C_y, C_z$ to layer $L_{i+1}$ results in two adjacent vertices being quasi-precoloured the same colour.

- There is no allowable pair for $L_{i-1}, L_i$ in which $L_i$ is assigned $C_x, C_y$, or there is no allowable pair for $L_{i+1}, L_{i+2}$ in which $L_{i+1}$ is assigned $C_y, C_z$.

- Assigning $C_x, C_y$ to layer $L_i$, and $C_y, C_z$ to layer $L_{i+1}$ quasi-bad forces a bad chain $P = [v_h...v_{i+1}]$.

The first two cases are straightforward, as in the base case.

In the third case, there is no colouring of $G$ that admits $\mathcal{P}$ such that the levels $L_{i-1}, L_i$ are assigned colours such that $L_i$ is assigned colours $C_x, C_y$, or there is no colouring of $G$ that admits $\mathcal{P}$ such that the levels $L_{i+1}, L_{i+2}$ are assigned colours such that $L_{i+1}$ is assigned colours $C_y C_z$. Therefore, there is no colouring of $G$ that admits $\mathcal{P}$ such that $L_i$ is assigned colours $C_x, C_y$ and $L_{i+1}$ is assigned colours $C_y, C_z$.

In the fourth case, by Lemma 8, the inductive hypothesis, and the quasi-bad forcing there is no colouring of $G$ that admits $\mathcal{P}$ such that $L_i$ is assigned colours $C_x, C_y$ and $L_{i+1}$ is assigned colours $C_y, C_z$.

Now assume that after executing Algorithm 2 $\mathcal{A}$ is not the allowable array of the input graph $G$, chain ordering $\mathcal{L}$ and list colouring $\mathcal{P}$.

Then there is a pair of colour assignments $C_x, C_y$ and $C_y, C_z$ to layers $L_i, L_{i+1}$ such that one of the following is true:

1. there is a vertex on $L_i$ precoloured by $\mathcal{P}$ with colour $C_z$,

2. there is a vertex on $L_{i+1}$ precoloured by $\mathcal{P}$ with colour $C_x$,

3. assigning these colours to these layers results in a quasi-precolouring such that two adjacent vertices are given the same colour,

4. there is no pair of colour assignments for $L_{i-1}, L_i$ that assigns $C_x, C_y$ to $L_i$.,

5. there is no pair of colour assignments for $L_{i+1}, L_{i+2}$ that assigns $C_x, C_z$ to $L_{i+1}$,

6. or the colour assignments $C_x, C_y$ and $C_y, C_z$ to layers $L_i, L_{i+1}$ are quasi-bad forcing.

We consider each of these cases. The first case would have been detected by the if statement on line 4 of Algorithm 3 after it was called by Algorithm 2, and so will not occur in $\mathcal{A}$ after processing, having been removed on line 6.

The second case would have been detected by the if statement on line 4 of Algorithm 3 after it was called by Algorithm 2, and so will not occur in $\mathcal{A}$ after processing, having been removed on line 7.

The third case would have been detected by the if statement on line 15 of Algorithm 3 after it was called by Algorithm 2, and so will not occur in $\mathcal{A}$ after processing, having been removed on line 16.

The fourth case would have been detected by the if statement on line 11 of Algorithm 2, and so will not occur, having been removed on line 12.

The fifth case would have been detected by the if statement on line 15 of Algorithm 2, and so will not occur, having been removed on line 16.

The sixth case would have been detected by the if statement on line 22 of Algorithm 2, and so will not occur, having been removed on line 23. □

By Lemma 10, Algorithm 2 gives the allowable array for $G = (V, E), \mathcal{L}, \mathcal{P}$. Each iteration except the last of the while loop on line 7 of Algorithm 2 removes at least one allowable pair. Because we have a fixed number (3) of colours and are selecting exactly two colours for each layer, there are at most $\binom{3}{2}! = 6$ allowable pairs for each pair of adjacent layers, and so at most $6n = O(n)$ total allowable pairs. Therefore, the while loop on line 7 executes $O(n)$ times. Because $O(n^2)$ work is done in each iteration, Algorithm 2 has time complexity $O(n^3)$.

## 4 Using the Allowable List to get a Colour Assignment

Once we have an allowable array $\mathcal{A}$, we need to extract a list of assignments of two colours to each layer that obeys $\mathcal{A}$. Let $Q$ be a colour assignment, $\mathcal{A}$ an allowable array for multichain ordering $\mathcal{L}$ of graph $G = (V, E)$.

Imagine that we have a good colour assignment $Q$ that obeys $\mathcal{A}$, and we decide to change the colour assignment in $Q$ at layer $L_i$ to some other colour pair that $\mathcal{A}$ allows for $L_i$. If we change only that layer, $Q$ may no longer obey $\mathcal{A}$, and may no longer be good. We informally give an algorithm for propagating a change through $Q$ to make $Q$ obedient and good.

We will first proceed by increasing level index. Then for a layer $j$ starting at $j = i+1$ we change the colour assignment $Q(j)$ to be $C_x, C_y$ for any $C_x, C_y$ such that if layer $Q(j-1)$ is $C_x, C_z$, then $[\{C_x, C_z\}, \{C_x, C_y\}]$ is in $\mathcal{A}[j-1]$. If

**Algorithm 2** Allowable Pair List Generating Algorithm Iterative. Input: Graph $G$, Layers $\mathcal{L}$, natural number *start*, list mapping $\mathcal{P}$

1: Array of list of pairs of pairs of size $|\mathcal{L}| - 1$ $\mathcal{A} \leftarrow null$
2: **for all** $i$ from 0 to $|\mathcal{A}| - 1$ **do**
3:    $\mathcal{A}[i] \leftarrow \{[\{C_1, C_2\}, \{C_1, C_3\}], [\{C_1, C_2\}, \{C_2, C_3\}], [\{C_2, C_3\}, \{C_1, C_2\}],$
   $[\{C_2, C_3\}, \{C_1, C_3\}], [\{C_1, C_3\}, \{C_1, C_2\}], [\{C_1, C_3\}, \{C_2, C_3\}], \}$
4: **end for**
5: Allowable Pair List Generating Algorithm Fixed($G, \mathcal{L}, start, \mathcal{P}, \mathcal{A}$)
6: changeMade $\leftarrow true$
7: **while** changeMade **do**
8:   **for all** $i$ from 0 to $|\mathcal{A}| - 1$ **do**
9:     **for all** Allowable pairs $W \in \mathcal{A}[i]$ **do**
10:       Let the allowable pair be $[\{C_x, C_y\}, \{C_y, C_z\}]$
11:       **if** there is no allowable pair $[*, \{C_x, C_y\}] \in \mathcal{A}[i-1]$ **then**
12:         Remove $[\{C_x, C_y\}, \{C_y, C_z\}]$ from $\mathcal{A}[i]$
13:         changeMade $\leftarrow true$
14:       **end if**
15:       **if** there is no allowable pair $[\{C_y, C_z\}, *] \in \mathcal{A}[i+1]$ **then**
16:         Remove $[\{C_x, C_y\}, \{C_y, C_z\}]$ from $\mathcal{A}[i]$
17:         changeMade $\leftarrow true$
18:       **end if**
19:     **end for**
20:     **for all** Allowable pairs $W \in \mathcal{A}[i]$ **do**
21:       COMMENT Here we're checking for quasi-bad forcing
22:       **if** forcesQuasiBad($G, \mathcal{L}, \mathcal{A}, i, W, \mathcal{P}$) **then**
23:         Remove $W$ from $\mathcal{A}[i]$
24:         changeMade $\leftarrow true$
25:       **end if**
26:     **end for**
27:   **end for**
28: **end while**

$\mathcal{A}[j]$ contains $[\{C_x, C_y\}, Q(j+1)]$ then we stop. Otherwise we continue on by incrementing $j$ and repeating the process for the next layer.

Then we can repeat this process starting at $j = i + 1$ and decrementing instead of incrementing $j$. We call this procedure the Colour Propagation Algorithm. To extract a colour assignment out of an allowable array instead of modifying one, we need only use this algorithm on a colour assignment with empty colour assignments and start at an arbitrary layer with an arbitrary colour assignment choice present in $\mathcal{A}$.

**Lemma 11.** *Let $G = (V, E)$ be a graph with multi-chain ordering $\mathcal{L}$, allowable array $\mathcal{A}$, and colour assignment $Q$ such that from layer $L_i$, $Q$ needs adjustment. If none of the entries in the allowable array are empty then the colour change propagation algorithm started at $L_i$ changes $Q$ to be a good colour assignment that obeys $\mathcal{A}$.*

*Proof.* We argue inductively in the case of propagation on decreasing level index $j$. Let $C_x, C_y$ be the colour assignment made by $Q$ to $L_{j+1}$. Then by the definition of allowable pairs, there is at least one allowable pair of colour assign-

**Algorithm 3** Allowable Pair List Generating Algorithm Fixed . Input: Graph $G$, Layers $\mathcal{L}$, natural number *start*, list mapping $\mathcal{P}$, list of allowed $\mathcal{A}$

1: COMMENT Remove pairs that are not allowable due to precolouring in $\mathcal{P}$
2: **for all** $i$ from 0 to $|\mathcal{A}| - 1$ **do**
3:    **for all** Colour $C_x \in C_1, C_2, C_3$ **do**
4:       **if** there is a vertex on $L_i$ precoloured with $C_x$ **then**
5:          Let $C_y, C_z$ be the two colours that are not $C_x$
6:          Remove $[\{C_y, C_z\}, *]$ from $\mathcal{A}[i]$
7:          Remove $[*, \{C_y, C_z\}]$ from $\mathcal{A}[i-1]$
8:       **end if**
9:    **end for**
10: **end for**
11: COMMENT Remove pairs that are not allowable due to quasi-precolouring forcing adjacent vertices of the same colour
12: **for all** $i$ from 0 to size of $\mathcal{A} - 1$ **do**
13:    **for all** Allowable pairs $W \in \mathcal{A}[i]$ **do**
14:       Let the allowable pair be $[\{C_x, C_y\}, \{C_y, C_z\}]$
15:       **if** there exist adjacent vertices $v_i, v_{i+1}$ in $L_i, L_{i+1}$ such that $\mathcal{P}[v_i] \cap \{C_x, C_y\} = C_y$ and $\mathcal{P}[v_{i+1}] \cap \{C_y, C_z\} = C_y$ **then**
16:          Remove $[\{C_x, C_y\}, \{C_y, C_z\}]$ from $\mathcal{A}[i]$
17:       **end if**
18:    **end for**
19: **end for**

ments to $L_j, L_{j+1}$ such that $C_x, C_y$ is assigned to $L_j$. Then we can make an assignment to $L_j$ that is obedient and good.

By the definition of an allowable array, any colour assignment that obeys an allowable array is good.

A similar proof applies in the case of iteration on increasing level index. □

## 5 Fixing Quasi-Bad Chains

We have shown how to build an allowable array, how to extract a good colouring assignment from an allowable array, and shown that there is a colouring of $G$ that admits a good colour assignment if and only if there is no quasi-bad chain.

We now show, given an allowable array with no empty entries and a good colour assignment obeying that array that has quasi-bad chains, how to produce a good colour assignment with reduced quasi-bad chains. Applying this inductively allows us to produce a good colour assignment with no quasi-bad chains, and therefore a 3-list colouring of $G$.

**Lemma 12.** *Let $\mathcal{A}$ be the allowable array for graph $G = (V, E)$ with multi-chain ordering $\mathcal{L}$ and list mapping $\mathcal{P}$. Let $Q$ be a good colour assignment to $\mathcal{L}$ such that $Q$ obeys $\mathcal{A}$, and $P = [v_h...v_l]$ is an N2 quasi-bad chain with respect to $Q$.*

*We can modify $Q$ such that it remains good, and all highest-indexed layers containing vertices of N2 quasi-bad chains have indices lower than $l$.*

**Algorithm 4** forcesQuasiBad: determines if input colour assignment pair is quasi-bad forcing with regard to a list of colour pairs $\mathcal{A}$. Input: Graph $G$, Layers $\mathcal{L}$, List of lists of pairs of colour assignments $\mathcal{A}$, integer $i$, colour assignment pair $W$, list mapping $\mathcal{P}$

---

Let $\{C_x, C_y\}$ assigned to layer $L_{i-1}$ and $\{C_y, C_z\}$ to layer $L_i$ be the colour assignments in $W$
List $Q \leftarrow \emptyset$
$Q(i-1) \leftarrow \{C_x, C_y\}$
$Q(i) \leftarrow \{C_y, C_z\}$
integer counter $\leftarrow i - 1$
boolean forced $\leftarrow$ true
set of colours belowColour $\leftarrow \{C_x, C_y\}$
**while** forced **do**
  **if** There is only one colour assignment pair $W'$ in $\mathcal{A}[counter - 1]$ that matches $[*, belowColour]$ **then**
    counter $\leftarrow$ counter $-1$
    counterColour $\leftarrow$ the first member of $W'$
    $Q(counter) \leftarrow$ counterColour
    belowColour $\leftarrow$ counterColour
  **else**
    forced $\leftarrow$ false
  **end if**
**end while**
**return** conservativeColouring($G$, $Q$, counter, $i$)

---

*Proof.* Let $P = [v_h...v_l]$ be a quasi-bad chain such that $L_l$, the layer that contains $v_l$ has the highest index of any quasi-bad chain.

Let $L_i$ where $h \leq i \leq l-2$ be the lowest-indexed layer that is adjustable. We know that there is such a layer because no colour assignment is quasi-bad forcing. Any quasi-bad forcing colour assignment would have been removed from the allowable array during its construction.

Without loss of generality, let $Q(L_i) = \{C_x, C_y\}$ and $Q(L_{i+1}) = \{C_x, C_z\}$. Then because all of $L_i, L_{i+1}, L_{i+2}$ contain vertices in a quasi-bad chain, then by Observation 7, $Q(L_{i+2}) = \{C_y, C_z\}$.

We modify $Q$ by setting $Q(L_i)$ to $\{C_y, C_z\}$ and propagating this in decreasing layer index using the Colour Assignment Propagation Algorithm. By the adjustability of $L_i$, this change obeys $\mathcal{A}$.

By Lemma 11 this produces a good colour assignment.

By Observation 7 there is no quasi-bad chain with respect to $Q$ that has an endpoint with a higher layer index than $i$ and an endpoint with a lower layer index than $i-1$ because all of $L_i, L_{i+1}, L_{i+2}$ are assigned colour $C_z$. Because colour assignments are the same in $Q$ below $L_{i+1}$ and $P$ was an N2 quasi-bad chain, the last vertex of every quasi-bad chain with respect to $Q'$ is in a layer with an index lower than $i$. □

**Lemma 13.** *Let $\mathcal{A}$ be the allowable array for graph $G = (V, E)$ with layers $\mathcal{L}$ and list mapping $\mathcal{P}$. We can produce $Q$, a good colour assignment to layers in $\mathcal{L}$ such that $Q$ obeys $\mathcal{A}$, such that there are no quasi-bad chains with respect to $Q$.*

*Proof.* By iterative application of Lemma 12. □

**Lemma 14.** *Let $\mathcal{A}$ be the allowable array for graph $G = (V, E)$ with multi-chain ordering $\mathcal{L}$ and list mapping $\mathcal{P}$. If there is at least one allowable pair in $\mathcal{A}$ for every adjacent pair of layers in $\mathcal{L}$, then we can produce a good colour assignment $Q$ of two colours to each layer in $\mathcal{L}$ such that the Conservative Colouring algorithm executed using $Q$ gives a 3-list colouring of $G$.*

*Proof.* Follows from Lemma 13 and Statement 9 □